\documentclass[
    oneside,
    fontsize=10pt,leqno]{article}
\usepackage[utf8x]{inputenc}
\usepackage[T1]{fontenc}
\usepackage[english]{babel}
\usepackage{enumitem}
\usepackage{amsfonts,amssymb,amsmath,amsthm}
\usepackage{graphicx}
\usepackage{tikz}
\usepackage{appendix}
\usepackage{natbib}
\usepackage{booktabs}

\long\def\thanks#1{\begingroup\def\thefootnote{
\fnsymbol{footnote}}\footnote[1]{#1}\endgroup}
\makeatletter
\def\maketitle{
   \null%
    \vskip 1cm
    {\centering{
    {\Large \@title\par}
    \vskip 0.4cm
    {\normalsize by}
   \vskip 0.3cm
    {\normalsize \@author}
    \vskip 0.9cm
    {\small \@date}
    \vskip 0.4cm}
             }}
\makeatother

\usepackage{geometry}
\usepackage{lscape}

\renewcommand{\normalsize}{\fontsize{10}{11.6}\selectfont}
\renewcommand{\footnotesize}{\fontsize{8}{9}\selectfont}

\makeatletter
\renewenvironment{quote}
               {\list{}{\footnotesize
                        \leftmargin=0.4cm
                        \rightmargin=0.4cm
                        \parsep=6pt}
                \item\relax}
               {\endlist}
\makeatother

\renewenvironment{abstract}{\begin{quote}}{\end{quote}}
\newcommand{\Keywords}[1]{\textit{Keywords:} #1}
\newcommand{\JEL}[1]{\textit{JEL classification code:} #1}

\usepackage{titlesec}
\titleformat*{\section}{\centering\normalsize\itshape}
\titlespacing*{\section}{0pt}{20pt}{6pt}
\titleformat*{\subsection}{\normalsize\itshape}
\titlespacing*{\subsection}{0pt}{12pt}{6pt}
\titleformat*{\subsubsection}{\normalsize\rmfamily}
\titlespacing*{\subsubsection}{0pt}{12pt}{6pt}
\titleformat*{\paragraph}{\normalsize\itshape}
\titlespacing*{\paragraph}{0pt}{8pt}{10pt}

\renewcommand{\appendix}{\par
  \pdfbookmark[1]{Appendix}{appendix}
  \setcounter{subsection}{0}
  \renewcommand{\thesubsection}{A.\arabic{subsection}}
  \setcounter{subsubsection}{0}
  \renewcommand{\thesubsubsection}{A.\arabic{subsection}.\arabic{subsubsection}}
  \setcounter{equation}{0} \renewcommand{\theequation}{A\arabic{equation}}
  \setcounter{table}{0}
  \renewcommand{\thetable}{A\arabic{table}}
  \setcounter{figure}{0}
  \renewcommand{\thefigure}{A\arabic{figure}}}

\setlist[itemize]{label=$-$,topsep=6pt,
parsep=0pt,itemsep=1pt,leftmargin=*,align=left}

\setlist[enumerate]{topsep=6pt,parsep=0pt,itemsep=1pt,
    leftmargin=*,align=left,widest=l)}

\usepackage[breaklinks,hidelinks]{hyperref}
\usepackage{hyphenat}
\makeatletter
\g@addto@macro\normalsize{%
  \setlength\abovedisplayskip{8pt}
  \setlength\belowdisplayskip{8pt}
  \setlength\abovedisplayshortskip{4pt}
  \setlength\belowdisplayshortskip{4pt}
}
\makeatother

\newtheoremstyle{JITEth}
{8pt}{8pt}{\itshape}{ }{\scshape}{ }{ }
{\thmname{#1} \thmnumber{\textsc{#2}}\thmnote{\ (#3)}}
\theoremstyle{JITEth}

\allowdisplaybreaks[3]

\usepackage{array}
\usepackage[font=footnotesize,labelfont=it,labelsep=newline,justification=centering]{caption}
\usepackage[para]{threeparttable}
\usepackage{multicol,multirow}
\usepackage{footnote}\makesavenoteenv{tabular}

\usepackage[center,footnotesize]{subfigure}
\usepackage[figuresright]{rotating}

\newenvironment{2Addresses}
{\bigbreak\parindent=0pt\setlength{\tabcolsep}{0pt}\small
\begin{tabular}{p{5.5cm}@{\hspace{1cm}} p{5.5cm}}}
{\end{tabular}}

\begin{document}

\title{Data Repurposing through Compatibility: \\A Computational Perspective}

\author{Asia J. Biega\thanks{
Asia J. Biega (corresponding author): Tenure-track Faculty, Max Planck Institute for Security and Privacy, Bochum, Germany. The author thanks the participants of the Institutional Economics Conference on The Machine Learning and the Law, in particular the paper commentators, Anthony Niblett and Michael Kurschilgen, for helpful feedback on the work. Thanks also go to Alessandro Fabris for useful comments on the paper draft, and to Lin Kyi, Franziska Roesner, and Cristiana Santos for helpful discussions on the formulations of data collection purposes.}}

\date{\today}
\maketitle

\begin{abstract}
Reuse of data in new contexts beyond the purposes for which it was originally collected has contributed to technological innovation and reducing the consent burden on data subjects. One of the legal mechanisms that makes such reuse possible is purpose compatibility assessment. In this paper, I offer an in-depth analysis of this mechanism through a computational lens. I moreover consider what should qualify as repurposing apart from using data for a completely new task, and argue that typical purpose formulations are an impediment to meaningful repurposing. Overall, the paper positions compatibility assessment as a constructive practice beyond an ineffective standard.


\JEL{C55, C80, D18, K24}
\end{abstract}

\section{Introduction}

The General Data Protection Regulation (GDPR)~\citep{GDPR2016a} 
is a legal framework governing the processing of user data in the European Union. Under the GDPR, various principles guide the handling of personal data, including the principle of purpose limitation. The principle mandates that data controllers process user data for specific, explicitly stated and legitimate purposes.

While the purpose limitation provision contributes to greater transparency and user control over their data, it also raises challenges for service providers seeking to rapidly innovate in a dynamic field like computing. Indeed, the possibility of reusing data in new contexts has contributed to progress in machine learning.
The GDPR, however, does offer mechanisms that allow for the repurposing of data, enabling service providers to navigate these challenges.

When a data controller intends to reuse data for a purpose beyond the initially specified purposes, they have two options: obtaining user consent or relying on compatible use. In their legal analysis, Seinen et al. highlight that while consent aligns more closely with the principles of privacy and user autonomy, the implementation of compatible use is often more practically feasible~\citep[p. 10]{seinen2018}.

Even though the implementation of compatible use might be simpler in practice, service providers nevertheless face the task of evaluating whether their intended data reuse falls within the boundaries of compatible use. Determining compatible use involves a comprehensive balancing test that considers five specific criteria. These criteria guide the decision-making process of the data controller and the documentation of that process. 

Seinen et al. further argue that compatibility assessments should not be a purely mathematical exercise~\citep[p. 4]{seinen2018}. While my analysis in this paper shows that they indeed cannot be fully automated, I argue that compatibility assessments can be supported by certain relevant computational evidence.

Moreover, in addition to exploring the difficulties associated with assessing compatible use, this paper investigates the ambiguity surrounding the definition of repurposing and suggests broadening the scope of compatibility assessments.

Lastly, acknowledging that the effective implementation of repurposing relies on purpose formulations, I examine the typical framing of purposes and its implications for compatibility assessments.

The paper contributes to the ongoing discourse around automation of data protection compliance, exploring how to turn compatibility assessments into a semi-automated, constructive practice. I also highlight the need for operational guidelines that would enable compatibility to be a true 'middle-way' between a de facto ban on data repurposing and an ineffective standard.

\section{Data Repurposing: a Background}

\subsection{Lawful initial processing and further processing}
To lawfully collect data from users under the GDPR, data controllers need to choose a legal basis for data processing. The GDPR specifies six such bases. In the context of online digital services, controllers typically rely on the legal bases of consent or legitimate interests.

Irrespective of the legal basis, service providers have to adhere to the principle of purpose limitation specified in Art. 5(1)(b) of the GDPR, which states that personal data shall be `collected for specified, explicit and legitimate purposes and not further processed in a manner that is incompatible with those purposes; [...]`. In practice, this means predefining the purposes for which data is being processed. 

Repurposing, or further processing, describes a situation where the service provider wishes to use data for ends other than those defined at the data collection time. Repurposing is thus related to data that has already been collected and in possession by data controllers. 

Controllers can use several legal mechanisms to repurpose data. Firstly, they can rely on data subject consent, explicitly specifying the new purpose. Secondly, they may rely on several exemptions specified in
Art. 5(1)(b) of the GDPR: ``further processing for archiving purposes in the public interest, scientific or historical research purposes or statistical purposes shall [...] not be considered to be incompatible with the initial purposes''. The interested reader may refer to \cite[pp. 8--9]{Finck_Biega_2021} for an examination of the challenges related to the interpretation of these exemptions. 
Finally, data controllers can determine if the new purpose is compatible with the purposes for which data has been previously processed using a 5-point \emph{compatibility assessment test}.

\subsection{Compatibility assessment}
The compatibility assessment test is a tool for determining whether two purposes of data processing are compatible. In conducting this test, a data controller must consider the following criteria, as specified in Art. 6(4) of the GDPR: 
\begin{enumerate}
\item[(a)] ``any link between the purposes for which the personal data have been collected and the purposes of the intended further processing;''
\item[(b)] ``the context in which the personal data have been collected, in particular regarding the relationship between data subjects and the controller;''
\item[(c)] ``the nature of the personal data, in particular whether special categories of personal data are processed, pursuant to Article 9, or whether personal data related to criminal convictions and offences are processed, pursuant to Article 10;''
\item[(d)] ``the possible consequences of the intended further processing for data subjects;''
\item[(e)] ``the existence of appropriate safeguards, which may include encryption or pseudonymisation.''
\end{enumerate}

In their legal analysis, Seinen et al. argue that a compatibility assessment is not a purely quantitative task, but should be a documented decision taken by a ``qualified legal or data protection expert''~\citep[p. 4]{seinen2018}. In this paper, I argue and show how this process could be supplemented with relevant computational evidence.

\subsection{Enforcement and guidelines}
Currently, there is a noticeable lack of specific computational and operational guidelines that could aid in the enforcement of compliant data repurposing. Moreover, it remains unclear whether engineering teams are actively using compatibility assessments in their day-to-day practice, as,  according to existing guidelines, these assessments should involve consultation with a data protection officer within the organization. The guidelines that do exist are very general in nature. For example, the directives of the French Data Protection Authority (DPA)~\citep{CNIL2022compatibility} simply allude to the necessity of conducting a compatibility test. Meanwhile, the UK DPA recommends utilizing an assessment based on their template for Legitimate Interest Assessment criteria~\citep{ICO2022basis}. However, a recent study by Kyi et al. has illuminated the challenges of determining when the legitimate interest legal basis can be used in practice~\citep{kyi2023legitimate}, raising a concern whether the method proposed by the UK DPA can truly enhance specificity in compatibility assessments. 

Considerably detailed, even if not computational, guidelines have been offered by the Article 29 Working Party~\citep[pp. 20--27]{WP203}. For instance, for the first compatibility criterion--the relationship between the purposes--it is suggested that ``This should not only be seen as a textual issue'', but that ``The focus should rather be on the substance of the relationship''~\citep[p. 23]{WP203}.

It is possible that the absence of detailed guidelines is a factor in the infrequency of purpose compatibility issues leading to penalties under the GDPR. Among the cases documented in The GDPR Enforcement Tracker,\footnote{\url{https://www.enforcementtracker.com/} accessed on Aug 31, 2023.} there's a single instance that cites Art. 6 (4),\footnote{\url{https://www.enforcementtracker.com/ETid-37} accessed on Aug 31, 2023.} pointing out that the data controller failed to perform the necessary compatibility assessment test altogether to validate their data processing. This situation underscores the potential advantages of integrating purpose compatibility assessments with computational practices to bolster enforcement measures.

\subsection{The appeal and challenges of purpose compatibility}
One of the primary concerns regarding purpose limitation revolves around its potential impact on innovation. Indeed, the possibility to repurpose data in new contexts has played a significant role in driving progress in machine learning. Therefore, data repurposing may play a crucial enabling role in fostering rapid technological innovation.

Seinen et al. highlight that while repurposing through consent aligns more closely with the principles of privacy and user autonomy, the practical implementation of compatible use is often more feasible in practice~\citep[p. 10]{seinen2018}. The current approach of collecting user consent through ubiquitous privacy notices has been considered ineffective and burdensome~\citep{habib2022consent}. Hence, repurposing through compatibility has the potential to additionally ease the burdens placed on individual data subjects in their day-to-day interactions with digital services. 

While data repurposing can undeniably bring about numerous benefits, it is important to recognize the potential harms that can emerge from such practices. A significant concern arises when data is repurposed without adequate scrutiny of the original and target contexts. For instance, practitioners might inadvertently optimize for inappropriate prediction proxies, or they could misconstrue the meaning of certain features within the data. It has been documented that biases can emerge when data is reused without a comprehensive understanding of the entire data generation process, including platform design and affordances, or platform behavioral norms~\citep[p. 10]{olteanu2019social}.

Compatibility assessments offer a promising middle ground that allows data controllers to dynamically evolve their services while still respecting user rights. However, while the problem of reuse compatibility has been examined from the legal~\citep{seinen2018, von2018principle} and ethical standpoints~\citep{Thylstrup2022reuse}, it has been notably underexplored from a computational lens. As a result, a multitude of issues may currently surface in practical settings. Among these concerns are questions about how to appropriately determine the compatibility of two purposes and  whether the ongoing practices resonate with and meet the expectations of data subjects.

In this paper, we delve into the examination of whether purpose compatibility assessments can be automated and if not, explore if computational evidence can be provided to support decision-making on purpose compatibility.

\section{Quantitative Aspects of Compatibility Criteria}
In this section, I examine the feasibility of automating compatibility assessment criteria, with a focus on identifying the aspects of these criteria that are most amenable to quantification. 

\subsection{Links between purposes}
The first criterion to be considered is ``any link between the purposes for which the personal data have been collected and the purposes of the intended further processing'' (Art. 6(4)(a) of the GDPR). 
%
Approaching this criterion from a functionality-based perspective, we can consider how closely related the functionalities described by these purposes are. Under a broad interpretation of 'related', we could consider any functionality that contributes toward the functioning of a system as compatible. In the case of recommender systems, for example, since using personal data for providing product recommendations, as well as for predicting the authenticity of product reviews, can be considered as necessary for effective functioning of a recommender system, the two purposes could be assessed as compatible. 

Alternatively, a narrower view of the link between purposes might require more parameters of a computational task to remain unchanged. For example, one could consider whether offering recommendations for new item genres after expanding the catalogue is compatible with offering recommendations for existing item genres.

Generally, it can be expected that the narrower the interpretation of the relationship between purposes is, the more feasible it becomes to measure functional purpose links quantitatively.

\subsection{Context of data collection and expectations of data subjects}
The second compatibility criterion requires us to consider ``the context in which the personal data have been collected, in particular regarding the relationship between data subjects and the controller'' (Art. 6(4)(b) of the GDPR). 
Context is a very broad concept, and one approach to grounding it is by referring to established theories such as contextual integrity~\citep{nissenbaum2009privacy}. Drawing upon this framework, compatibility assessments could consider changes in contexts (domains), or in the parameters of information flows (data subject, sender, recipient, information type, and transmission principle).

Reasonable expectations of data subjects could be established through user studies. However, I would contend that the objective of such studies should not be to determine user perceptions of the relationship between two purposes, as previous studies have consistently shown that the functioning of digital systems might not be sufficiently transparent to users~\citep[e.g.][]{rader2015understanding, eslami2016first, ngo2020exploring}. Instead, the focus of the study could be on evaluating the level of perceived system behavior change when data is used for the new purpose, such as a noticeable improvement or a degradation in performance, the perception of emergent system capabilities~\citep{emergentCapabilities}, or qualitative changes in the results received.

While initially, the design of user studies for purpose compatibility assessments would need to be tailored to specific use cases, one might envisage the development of a validated instrument that could be applied across different systems and purposes over time.

\subsection{The nature of personal data}
The third compatibility criterion is ``the nature of the personal data, in particular whether special categories of personal data are processed, pursuant to Article 9, or whether personal data related to criminal convictions and offences are processed, pursuant to Article 10'' (Art. 6(4)(c) of the GDPR). Art. 9(1) of the GDPR defines special categories of data as ``personal data revealing racial or ethnic origin, political opinions, religious or philosophical beliefs, or trade union membership, [...] genetic data, biometric data, [...] data concerning health or data concerning a natural person's sex life or sexual orientation''.

While a straightforward interpretation of this criterion might simply look at whether special categories of data are explicitly present in the processed datasets, a more careful interpretation should consider whether special categories of data can be \emph{predicted} from the processed data. Or more specifically, in the context of repurposing, whether such predictions become more accurate under the processing for a new purpose. Prior work shows, for example, that it is possible to predict sexual orientations from movie rating data~\citep{narayananDeanonymization}. Evidence for this criterion in compatibility assessments could be generated by predictors for each special category of data, assuming that high-quality ground-truth data were available. Such data could be obtained through data donations or be voluntarily self-reported by data subjects

\subsection{Consequences of data processing}
The fourth compatibility criterion to be taken into account are ``the possible consequences of the intended further processing for data subjects'' (Art. 6(4)(d) of the GDPR). Ethical, safety, and social implications of data processing are especially significant in this regard. Anticipating such consequences is a complex task, as evidenced by research fields focusing on these problems, including the interdisciplinary community of the Fairness, Accountability, and Transparency (FAccT) conference,\footnote{\url{https://facctconference.org/}, accessed on Aug 23, 2023.} or by the need for guidelines for computer scientists on how to develop societal impacts sections in their papers~\citep{boyarskaya2020overcoming}. 
Practically, techniques such as ethical speculation~\citep{ethicalSpeculation}, or construction of hypothetical technology scenarios~\citep{mhaidliFiction} based on the technique of design fiction~\citep{designFiction}, might prove particularly useful for assessing repurposing that may lead to emergent capabilities of a system. 

\subsection{Existence of appropriate safeguards}
The final compatibility criterion is ``the existence of appropriate safeguards, which may include encryption or pseudonymisation'' (Art. 6(4)(e) of the GDPR). While this criterion is not particularly amenable to a quantitative interpretation, one could measure the mechanism-dependent efficacy of selected approaches in safeguarding users. For instance, if data is pseudonymized, we could measure risks of reidentification of individuals under a selected adversary model. 

The criterion could moreover be seen as a means for ameliorating the negative impacts identified by other criteria and thus reducing risks for data subjects. Beyond encryption and pseudonymization, a variety of other techniques could be considered here, including but not limited to, data perturbation~\citep{nissenbaum2009trackmenot,Kargupta2003Perturbation}, randomly mixing user profiles~\citep{rebollo2012exchange, biega2017solidarity,eslami2017Privacy}, or construction of synthetic datasets~\citep{vreeken2007Privacy, ping2017datasynthesizer}.

\subsection{Interim conclusions}
Summarizing this section, I conclude that the compatibility assessment criteria are not amenable to full automation, although, as preliminarily highlighted, certain aspects of each criterion could be quantified. Thus, in this paper I do not argue for turning compatibility assessments into a purely computational task. Instead, the quantified aspects of compatibility could form auxiliary evidence for the repurposing decision-maker. In the next section, we delve into a case study of repurposing to examine some of the caveats of quantitative purpose link measurements.

\section{A Case Study of Repurposing}
We explore the feasibility and issues with automated measurement of links between purposes, grounding our arguments in a case study of recommender systems. While repurposing could take many forms, here we examine the following setting: A service provider adds a new category of items to their catalogue. Is using item ratings of items already existing in the catalogue for the purpose of providing recommendations of the new items compatible with the purpose of providing recommendations of the old items? Relevant scenarios include movie and music recommendations on streaming platforms, or content recommendations on social media.

\subsection{Instantiating data repurposing}
The concrete instantiation of the data repurposing problem we examine in this study is as follows. A data controller is lawfully processing user data for the purpose of providing recommendations of items (for instance, movies). Items belong to certain categories $G = {g_1, ..., g_k}$ (in the context of movie recommendations, those could be movie genres, like drama, romance, comedy, etc.) 

For each user $u_i$, the personal data that is being processed consists of a set of user ratings over items $i_j$: ${r_{i,j}}$. Ratings are typically expressed on a discrete scale, binary (thumbs up -- thumbs down) or more fine-grained (for example, 1--5). This data is observed, that is, explicitly provided by the data subjects. 

The volume of this data (typically, up to several hundreds of ratings per user) is much smaller than the number of items in the catalogue. The goal of the service is to provide recommendations of items that the user has not rated (thus, presumably has also not seen): using observable user ratings $r_{i,j}$, the controller makes predictions about unseen ratings, $r'_{i,j}$. Ideally, these recommendations should be of high quality, that is, likely to lead to high user satisfaction. Recommender systems use observable item ratings to infer user preferences, in particular combining them with data of other users in various ways. For instance, a technique of \emph{collaborative filtering}~\citep{goldberg1992cf}, is based on the assumption that similar users will like similar items.

The controller adds new item genre $g_{k+1}$ to the catalogue, and wants to provide recommendations over these items to their users. The question we want to examine here is: can we use existing user ratings for items from genres $G$ to provide recommendations of items from genre $g_{k+1}$ based on purpose compatibility? Is the purpose of providing recommendations of movies from genres $G$ compatible with the purpose of providing recommendations of movies from genres $G \cup g_{k+1}$? The remainder of this section seeks to surface some of the intricacies of this question.

\subsection{Data}

In our experiments, we focus on the domain of movie recommendations. We use one of the standard datasets from this domain, the MovieLens-100k dataset~\citep{movielens100k}. The dataset consists of 100,000 ratings of value 1-5 from 943 users on 1682 movies. Each user has rated at least 20 movies. While the dataset is relatively old, its unique characteristic is that it contains some demographic information about the users (age, gender, occupation, zip code). Each movie may belong to more than one genre; overall, there are 18 different genres represented in the dataset.

\subsection{Measuring links between purposes}
We begin with the first criterion of compatibility assessment: ``any link between two purposes''. How might we measure links between two purposes of providing recommendations over different sets of items? Here, we explore two ways to measure such links: one using exclusively data about items and genres, and another using (subsets) of existing personal user data.

\subsubsection{A measure based on non-personal data}
We first show how to measure links between two purposes without personal user data. In our case scenario, we can compute the similarity between two movie genres using the \emph{movie-genre} annotations. Since each movie may belong to more than one genre, we can reason about the compatibility of purposes by quantifying movie overlaps between genres. 

Overlaps between genres can be measured, for instance, as conditional probabilities of a movie belonging to genre $g_y$ given that it belongs to genre $g_x$: $P(i\in g_y|i \in g_x)$.\footnote{Conditional probabilities are not the only measure that could be applied here. An alternative might be, for instance, the Jaccard distance, which is a (symmetric) measure of overlap between two sets.} In our dataset, this quantity can be computed as $ {|\{i : i\in g_x \land i\in g_y \}|} \over {|\{ i : i \in g_x \}|}$.  Note that the measure is not symmetric. Here we might use the new genre in the condition, capturing the following intuition: If I take a movie from the new genre, how likely is it that I have already been providing recommendations for it as part of the existing genre recommendations?

The measure will give us a value between 0 and 1 for a particular pair of categories (one to be added and one existing). How should a measure like this be used for a compatibility assessment? Here we stumble upon the first issue with automated compatibility assessments: When a continuous measure is to be converted into a binary decision (compatible or not compatible), we need to pick a value threshold. However, without further research into the relationship between the value of such purpose link metrics and legal compatibility, it is unclear how this  threshold should be established. 

\begin{figure}[h]
\centering
\includegraphics[width=0.9\textwidth]{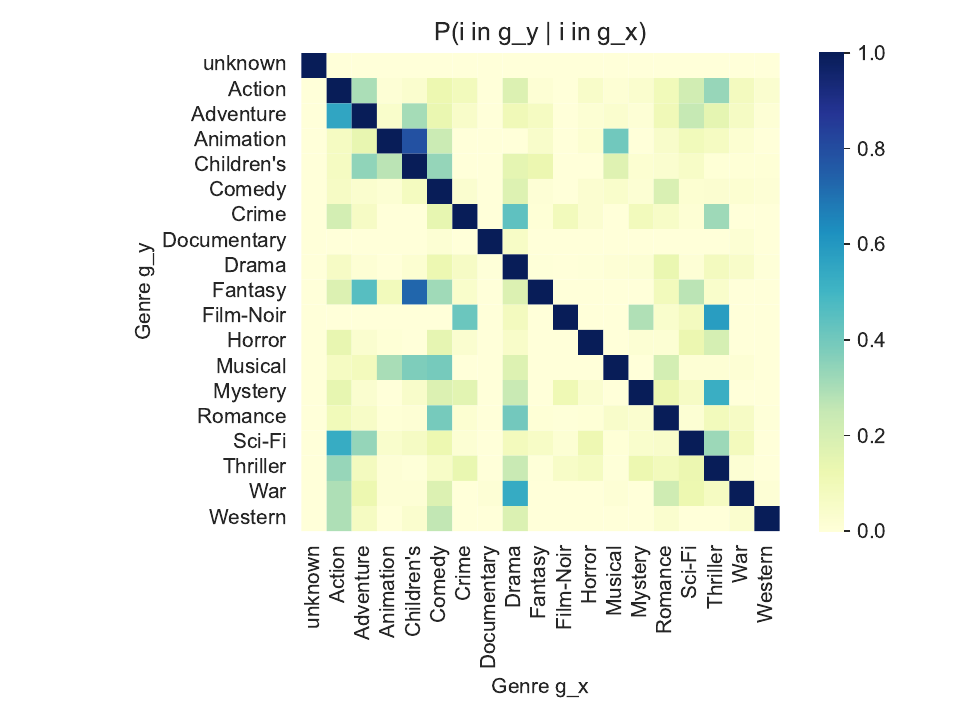}
\caption{Conditional probabilities of a movie belonging to genre $g_y$ given that the movie belongs to genre $g_x$, as estimated from the MovieLens-100k dataset. We explore the measure in the context of data repurposing: is providing recommendations for genre $g_x$ (a new genre to be added to the catalogue) compatible with providing recommendations for genre $g_y$ (an existing genre)? The conditional probability measure captures the following intuition: if I add movies from genre $g_x$ to the catalogue, how likely is it I have already been providing recommendations for these movies as part of genre $g_y$?}
\label{fig:link-non-personal}
\end{figure}

What we can do in this analysis, however, is to examine the genre pairs with the highest value of the metric. Figure~\ref{fig:link-non-personal} presents the results computed over the MovieLens-100k dataset. Irrespective of the selected threshold, according to the conditional probability measure, a dark square in the heatmap suggests that the respective genre on the x-axis might be most admissible in the compatibility assessment, given that we already have been providing recommendations for the given genre on the y-axis. 

Let us examine a few examples of genre pairs with the highest value of the metric. Several such cases provide a sanity check for the idea of computing purpose links this way. For instance, we could consider adding $g_x = \text{Action}$ if we already provide recommendations for $g_y = \text{Adventure}$ (conditional probability of $0.56$), or $g_y = \text{Sci-Fi}$ (conditional probability of $0.53$). Similarly, we could consider adding $g_x = \text{Comedy}$ if we already provide recommendations for $g_y = \text{Romance}$ (conditional probability of $0.39$), or $g_y = \text{Musical}$ (conditional probability of $0.39$). 

Several examples, however, surface issues with this approach. The measure suggests we could add $g_x = \text{Children's}$, if we are providing recommendations for $g_y = \text{Animation}$ (conditional probability of $0.79$), or $g_y = \text{Fantasy}$ (conditional probability of $0.73$). We could perhaps expect some adults without children to be surprised to start receiving recommendations for children's movies. Even more problematically, we get a relatively high value of the measure ($0.34$) suggesting we could start providing $g_x = \text{Adventure}$ or $g_x = \text{Comedy}$ recommendations if we are providing recommendations for $g_y = \text{Children's}$ movies. Since movies from other genres might be inadequate for children, other assessment criteria (e.g., the feasibility of implementing a safeguard preventing new items from being recommended to minors) might be necessary.

\subsubsection{A measure based on personal data}
The purpose link measure discussed in the previous section relies on two assumptions: (i) the availability of item category annotations, and (ii) the possibility for items to belong to multiple categories. However, it is important to acknowledge that these assumptions may not always hold. Consequently, we explore an alternative measure that utilizes user-generated data, specifically item ratings. While this alternative measure introduces a different perspective on purpose links, it does not completely eliminate assumptions. In this case, we assume that the data controller is able to lawfully process the data for the purpose of establishing compatibility. For instance, one plausible scenario involves obtaining consent from a sample of users to determine compatibility and subsequently processing the data for the purpose of providing new category recommendations to the remaining users if compatibility is established.

Assuming that we can indeed leverage user-generated data, we can measure whether there is a positive correlation between user preferences for movies from an existing genre and their preferences for movies from the new genre. 
The underlying intuition behind this measure is that similarity between two genres can be established by population-level preference trends.

We measure the link between two genres as follows. First, for each user, we average the user's ratings for each genre. Then, for each pair of genres, we compute the Spearman rank correlation $\rho$ between the averaged user ratings. 

\begin{figure}[h]
\centering
\includegraphics[width=0.9\textwidth]{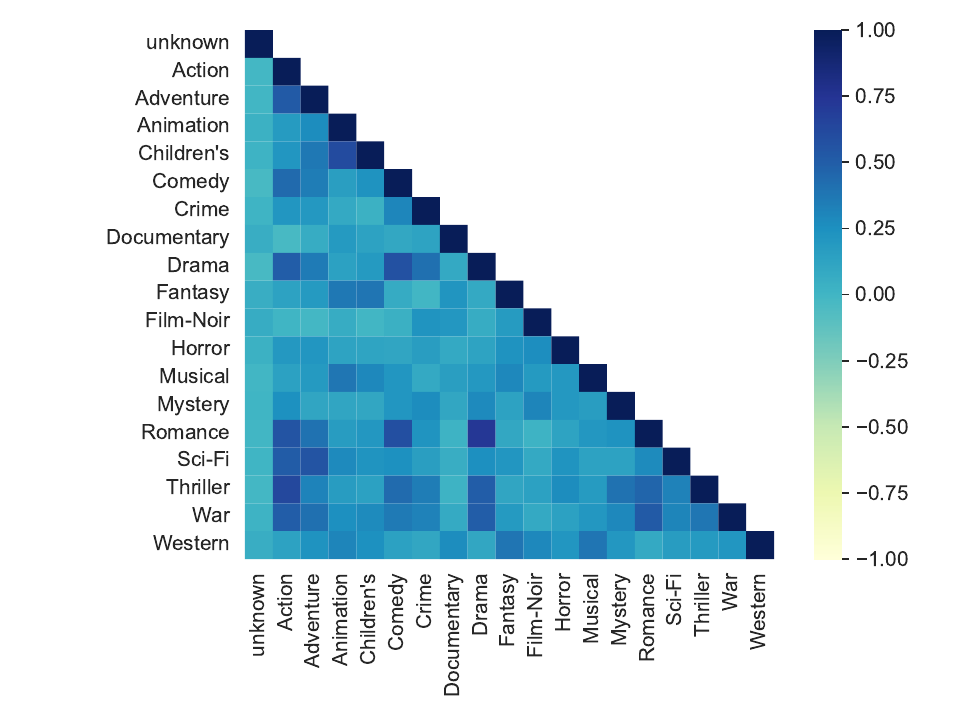}
\caption{Spearman rank correlations between average user ratings of movies belonging to genre $g_x$ and average user ratings of movies belonging to genre $g_y$. A high value indicates a positive correlation between user ratings of movies in $g_x$ and movies in $g_y$. The measure captures the similarity between two genres in terms of population-level movie preferences.}
\label{fig:link-personal}
\end{figure}

Figure~\ref{fig:link-personal} presents the results. As with the previous measurement method, we can observe strong similarity between genre pairs like Drama and Romance ($\rho=0.73$, $p<0.01$), or Action and Thriller ($\rho=0.63$, $p<0.01$). 

Note this method of measurement would not fully address the issue with Children's category being unlike other genres, with the highest correlation of $\rho=0.61$, $p<0.01$ between the Children's and Animation categories. Without very conservative thresholding, one would need to implement appropriate additional safeguards when expanding the catalogue for users receiving recommendations for children's movies. 

The appeal of a purpose link measurement method based on user item ratings lies in capturing aspects of user expectations. A pitfall, however, is that such observational data is often biased. For example, since some users tend to rate movies higher on average than others, recommendation algorithms need to debias data accordingly~\citep[pp. 13--15]{Cacheda2011tedency}. This bias limits the interpretation of our analysis. Even though the majority of correlations, except for a few cases, are significant (after Bonferroni corrections for multiple hypotheses), user rating tendencies as well as other domain-relevant biases would need to be accounted for as confounding factors.

\subsection{Impacts on the experience and expectations of data subjects}
\label{sec:user-expectations}
While we have demonstrated that, to a certain extent, it is possible to establish quantitative connections between purposes, a question remains of what changes from the data subject's perspective when their data is used for recommendations in a new genre. We can anticipate at least three kinds of change.

Firstly, the most apparent change is that users will begin receiving recommendations of a different nature. Depending on the specific genre or items involved, this change may go unnoticed by the user (for example, they may not realize that a war category has been added when a drama category already existed), or they may be surprised by unexpected recommendations. The latter scenario could be particularly critical if the new genre is sensitive in nature (such as movies about religion or other personal topics), vastly different from existing genres (like adding a horror genre to a catalog of children's movies), or if it triggers negative associations for the user (for instance, a war veteran with PTSD being recommended war movies).

Secondly, the composition of recommendation results may change. Typically, recommendations are presented to users in a top-k fashion, where ratings are predicted for all items in the catalog and the system selects the ones most likely to be enjoyed by the user. In this case, users may be dissatisfied if they receive fewer recommendations from a category they like, as the top-k list becomes partially filled with recommendations from the new genre.

Finally, the quality of predictions may be affected. Often, when a new category of items is introduced, the entire recommendation model undergoes a change. A single model is responsible for making predictions across all items and genres for all users. Consequently, the predictions for existing items may be altered, potentially leading to an increase or decrease in prediction errors. It is worth considering whether this paradigm of sudden change should be abandoned in favor of more gradual model updates~\citep{Patro_Chakraborty_Ganguly_Gummadi_2020}. 

Even if in the process of task update the prediction quality improves, users may feel unsettled by the sudden exceptional performance of the system -- a phenomenon termed the ``Uncanny Valley effect''~\citep[p. 5]{Bernardi2019Successful}. Although this is unlikely to be the case in our simplified case study, it is a pertinent consideration for many types of computational models, including the recently improved chatbots based on large language models.

The extent to which the system's performance aligns with user expectations is an active area of study. However, to the best of my knowledge, no study thus far examined user expectations of model performance in the context data processing purposes. Thus, it is currently challenging to assess whether repurposing would violate user expectations, or whether these expectations are clearly defined in the first place.

\subsection{Feasibility of generalizing purpose link measurements}

Do the approaches for measuring links between purposes easily generalize to other scenarios? 
When considering scenarios that fall within the same definition of data processing purposes as the ones we consider in this paper (providing recommendations within a certain category of items), the measurement approaches we have investigated in the present paper can generally be applied. In such cases, utilizing a non-personal-data-based approach remains viable as long as we have category annotations for items at our disposal and items may belong to multiple categories, such that we can measure overlaps between categories using this data. However, generalization may not be possible if the new items introduced to the catalogue are fundamentally different in nature. For example, transitioning from offering movie recommendations to providing recommendations for hiking gear may require a different approach as these categories do not share any items. The personal-data-based approach could still be employed, however, if it can be assumed that individuals who share similar movie preferences are likely to have similar preferences for hiking gear.

It is important to note that the measurement approaches we have  examined in this paper do not generalize to scenarios where the computational task significantly changes. For instance, if the purpose shifts from offering item recommendations to predicting other user characteristics, such as age, purpose links cannot easily be measured based on item or user data. In such cases, the development of new approaches specifically tailored to the changed task becomes necessary, and if that is not possible, it should perhaps be assumed that the purposes are not compatible.

Furthermore, the need for a different approach becomes more pronounced if the computational task at hand is different from what we have studied in this paper. For instance, consider a classification model designed to predict income. Repurposing the data for a different predictive task (such as predicting age), would require us to tailor purpose link measurement to the intricacies of a classification task, including determining what it means for two target variables to be related. 

Overall, the measurement of links between purposes appears to be highly context-dependent. It may necessitate human involvement in determining the appropriate computational methods and require further research to establish a comprehensive catalog of approaches suitable for different contexts.

\subsubsection{Interim conclusions}
Our study suggests that computing could offer certain quantitative evidence for compatibility assessments, yet the assessment process itself is not particularly amenable to automation:
\begin{itemize}
\item Establishing quantitative links between purposes necessitates domain knowledge to determine the relevant measures within a specific context. Additionally, the interpretation of these measures may involve human-in-the-loop steps, such as setting thresholds to determine compatibility based on metric values.
\item It is essential to establish clearer connections between purpose link metrics and data subject expectations. In particular, the understanding of data subject expectations with regard to data processing purposes in fine-grained computing scenarios requires further research.
\item The definition and operationalization of repurposing can vary across different computing domains. While this paper focuses on scenarios like offering recommendations for new items within a new category, there are other possibilities where the purpose changes are more significant. Quantitatively measuring links between purposes becomes even more challenging in such cases, necessitating case-specific and domain-informed examinations. Moreover, the ambiguity surrounding the operationalizations of repurposing raises the question of what should be considered as repurposing in the first place. This question will be explored in the subsequent section.
\end{itemize} 

\section{What qualifies as repurposing}
A question I would now like to consider is when a compatibility assessment test should be conducted in the first place. Specifically, we need to determine what qualifies as repurposing and when we can assume that data continues to be processed for its original purpose.  While certain scenarios, such as a change in computational task from recommendation to classification, or switching to predictions for different target variables, can easily be identified as repurposing, I argue that various data- and algorithm-related changes in a computing system should also be considered as repurposing, even if the computational task itself remains unchanged. These changes should then undergo a compatibility assessment test to ensure their compatibility with the original purpose.

The GDPR does not explicitly define repurposing. However, Article 5(1)(b) implicitly links the definition to the original purpose of data processing. It states that personal data should be \emph{``collected for specified, explicit and legitimate purposes and not further processed in a manner that is incompatible with those purposes''}. We will return to the issue of purpose formulation in the subsequent section of this paper.
Instead, in this section, I will examine several data- and algorithm-related changes in a system which could trigger a negative assessment based on reasonable user expectations and the processing of special categories of data, should the compatibility assessment test be applied.

\subsection{Algorithm changes}
First, let us consider a scenario involving an algorithm change -- the computational task remains the same, but the method of computation is altered. In the context of recommender systems, this could entail, for example, transitioning from a matrix-factorization-based approach to a deep-learning-based approach. Service providers often update their algorithms to stay current with advancements in the field, especially if the update promises enhanced performance or improved user satisfaction. However, as we discussed in Section~\ref{sec:user-expectations}, it is important to recognize that changing performance might lead to various issues that violate user expectations. In cases of performance improvement, these issues can range from users being surprised by the composition of the results to feeling unsettled by the system's uncanny predictive accuracy or uneasiness about the system's emergent capabilities. User expectations may also be violated by a decreased system performance. For instance, users might not be willing anymore to allow for processing of their data in exchange for a worse service.  

Additionally, beyond localized algorithmic changes, changes to platform values or policies may also lead some users to withdraw their consent data processing, or leave the platform completely. Recently, this process has been exemplified when a number of users migrated to Mastodon triggered by functionality changes that followed Twitter's ownership transfer~\citep{jeong2023exploring}.

\subsection{Data distribution shifts}
Users may encounter the aforementioned changes in system behavior when there is a global shift in the characteristics of user data, often referred to as \textit{data distribution shift}~\citep{quinonero2008dataset}. This shift can occur, for example, when a service experiences a significant increase in the number of users from a specific background. In the context of recommendations, for instance, when users from a particular demographic that exhibits high interest in violent movies join a platform, the algorithm might end up increasing the recommendation rate of such content to existing users due to popularity bias~\citep{bellogin2017statistical} -- a recommendation phenomenon where popular items are recommended more frequently.

There are various events that can trigger a data distribution shift. Firstly, as illustrated in the previous example, the user base of an online service may change when users join or migrate to another service in large numbers. Targeted advertising can attract users from specific demographics to a service, resulting in such shifts. However, a similar change can occur even if the user base remains unchanged but users delete certain content from their profiles. This deletion could be prompted by a public privacy controversy or a request from the platform to reduce storage volume.

Moreover, it has been observed that predictions about inactive users (users who do not generate new data) can become more accurate over time due to data contributions from other active users~\citep{Rizoiu2016evolution}. In this case, the data distribution might not necessarily shift, but the accuracy of estimates improves. An issue arising here is not only the potential violation of user expectations but also the sudden possibility of processing sensitive data. Some attributes considered by~\cite{Rizoiu2016evolution} indeed include special categories of data, such as gender.

\subsection{Data merging}
Purpose formulations often lack detailed specifications regarding the types of user data to be used for specific purposes. As a result, another relevant scenario to consider is \emph{data merging}, where two sets of data points pertaining to a data subject, initially collected for separate purposes, are combined. These sets could include user self-reported attributes and behavioral data gathered from the platform. The objective of data merging might involve enhancing the service in relation to one of the original purposes. Similar to the issues discussed in the previous section, besides the concerns regarding system performance and emergent capabilities, it is important to consider the potential transformation into special categories of data in this context as well. Prior foundational research in privacy has demonstrated that combinations of individually non-identifying attributes (such as a zip code and a date of birth) can lead to the creation of unique personal identifiers. This observation forms the basis for privacy models like k-anonymity~\citep{kAnonymity}. 

\subsection{Toward continuous compatibility assessments}
\label{sec:continuous-assessments}
Through the above analysis, several common themes have emerged. Changes in the system may mean that data subject expectations of system behavior can be violated, leading to feelings of surprise or unsettlement. These violations can occur not only when the computational task changes but also when there are changes in the algorithm, data properties, or user base. Consequently, it can be argued that compatibility assessments should not be one-time procedures triggered solely by significant task changes. Instead, they should be an ongoing process of monitoring and re-evaluating compatibility. 

Metrics of interest for such assessments could include system performance, user satisfaction, user surprise, prediction accuracy for special categories of data, performance on tasks beyond the explicitly defined scope to anticipate emergent capabilities, and more. While these metrics are commonly used to evaluate system performance, they are not commonly associated with compatibility assessments.

Significant changes in the tracked metrics could prompt a comprehensive compatibility assessment conducted by a data protection specialist. They may also warrant an investigation by software engineers to determine if new technical safeguards are necessary. Additionally, an interface-level prompt could be presented to users, informing them about the current state and behavior of the system, with a clear option to withdraw their data processing consent if desired. 

\section{Repurposing and Purpose Formulations}
In the previous sections, we have been reasoning about repurposing using a specific formulation of purposes -- providing automated recommendations for items of a specific category. This modeling choice reflects the assumption that purposes are formulated meaningfully and in a detailed manner.  The final consideration I want to make in this paper, is whether current, typical formulations of purposes live up to this expectation.

First, let us consider purpose formulations for cookie data commonly used in the context of online advertising. TCF (The Transparency and Consent Framework) is ``an accountability tool that relies on standardisation to facilitate compliance with certain provisions of the ePrivacy Directive and the GDPR" and ``a voluntary standard intended for use by three categories of stakeholders"~\citep{tcf}, including publishers, vendors and content management platforms. TCF de facto standardizes data processing purposes for web browsing users, and version 2.2 of the framework proposes the following set of purposes~\citep{tcf-purposes}: 
\begin{itemize}
\item Store and/or access information on a device,
\item Use limited data to select advertising,
\item Create profiles for personalised advertising,
\item Use profiles to select personalised advertising,
\item Create profiles to personalise content,
\item Use profiles to select personalised content,
\item Measure advertising performance,
\item Measure content performance,
\item Understand audiences through statistics or combinations of data from different sources,
\item Develop and improve services,
\item Use limited data to select content,
\item Ensure security, prevent and detect fraud, and fix errors,
\item Deliver and present advertising and content,
\item Match and combine data from other data sources,
\item Link different devices,
\item Identify devices based on information transmitted automatically,
\item Use precise geolocation data,
\item Actively scan device characteristics for identification.
\end{itemize}

Out of the above purposes, few are pertinent for recommendations: ``Create profiles to personalise content'', ``Use profiles to select personalised content'', ``Measure content performance'', ``Develop and improve services'', ``Use limited data to select content'', ``Deliver and present advertising and content''. However, in the empirical section of this paper, we focused on automated recommendations, where data is usually gathered from authenticated users. Therefore, let us also investigate how purposes are formulated in this scenario. The following list presents the data processing purposes as specified for a Google account user:\footnote{\url{https://policies.google.com/privacy?hl=en-US\#whycollect}, accessed Aug 31, 2023.}
\begin{itemize}
\item Provide our services,
\item Maintain \& improve our services,
\item Develop new services,
\item Provide personalized services, including content and ads,
\item Measure performance,
\item Communicate with you,
\item Protect Google, our users, and the public.
\end{itemize}

Both of these purpose lists expose two significant issues with the current purpose formulations. Firstly, purposes often combine numerous functionalities into a single statement. For instance, a broad interpretation of ``Provide our service'' might encompass data collection for every computational task performed by the data controller. Even without such generous interpretations, purposes like ``Provide personalized services, including content and ads'' clearly demonstrate the use of conjunctions to describe functionalities that utilize similar input personal data but rely on different non-personal data (ads or content). These functionalities are typically implemented as separate system components employing distinct computational techniques. Conversely, TFC purposes divide personalized content-related purposes into a detailed list of functionalities, which creates an additional burden of control over functionalities that are typically executed together (e.g., ``Create profiles to personalize content'' and ``Use profiles to personalize content'').

The second concern regarding the phrasing of purposes is that they are not readily amenable to a quantitative interpretation. Some purposes, such as ``Provide our services'' or ``Develop new services'', do not specify any particular computational tasks. Purposes like ``Measure performance'' would require additional details regarding the object of performance or the metrics applied. Indeed, more specific purposes would offer greater possibilities for implementing data protection laws: Shanmugam et al. and Biega et al., for instance, demonstrate potential interpretations of the data minimization principle that would be feasible with purpose formulations tied to concrete performance metrics~\citep{shanmugam2022dmin, biega2020operationalizing}.

Current purpose formulations are too broad to allow for meaningful repurposing. On the other hand, overly specific purposes might lead to a situation where compatibility can never be achieved in practice, limiting the benefits of repurposing for technological innovation. Moreover, overly technical purpose descriptions might make it harder for data subjects to give meaningful consent. Finding the right formulations will be a balancing act.

One approach to consider in the context of repurposing might be an appropriate version of a ``layered notice'', as suggested by Article 29 Working Party for data processing notices~\citep[p. 16]{WP203}. Here, one layer could contain purpose descriptions developed for collecting user consent, while another might contain descriptions developed for determining purpose compatibility, potentially in a more technical manner geared toward computing practitioners.

Technical approaches to purpose formulation using specially designed formal languages have already been explored in the context of automation of compliance ~\citep{Karami2022dpl}, or in the context of privacy policies~\citep[e.g.,][]{cranor2002platform}. Descriptions expressed in formal languages~\citep[pp. 14-18]{koops2011flexibility} or through ontologies~\citep[p. 6]{fouad:hal-02567022} might indeed allow for formal reasoning about relationships between purposes.

Much is moreover to be learned from the domain of cookie data in terms of issues with purpose formulations. For instance, many cookie expressions have been shown to violate the GDPR requirements~\citep{santos2021violations}, including specificity. It remains to be seen whether the impossibility to meaningfully repurpose data may count as a sign of lack of purpose specificity as well.

In summary, although our analysis in this paper suggests that certain quantitative evidence could be valuable for purpose compatibility assessments, the current use of generic and ambiguous purpose formulations renders such evidence unnecessary. Repurposing can only be meaningfully carried out if purposes are initially formulated in a meaningful manner.

\section{Alternative Perspectives}

\subsection{From functionality to harms and benefits}
This paper adopts a distinctly functionality-based perspective of purpose compatibility assessments. We evaluate compatibility by measuring correlations of the changing system outputs, and potential breaches of user expectations regarding system behavior. This perspective could be enriched by considering various other facets, such as an examination of benefits, harms, or the value derived by the data subject from additional data processing. Notably, several compatibility criteria, such as the consequences of data processing or the existence of appropriate safeguards, directly emphasize these aspects. Risk-based interpretations of compatibility have also been examined in the legal literature before~\citep{custers2016big,von2018principle}.

\subsection{User intents}
Rather than solely relying on consent for specific data processing objectives, we can envisage an alternative approach that gives users greater control. In this approach, users could articulate their own purposes for using the system and express their expectations for how their data is processed. These expectations can then be examined in light of the intended data usage. Whenever a service provider considers repurposing the data, they should assess if the new use aligns with the original intent of the user. This nuanced approach to consent may take into account additional contextual information. For example, reusing data for research might be viewed differently depending on whether the research is for commercial gain or serves the public interest. Such an approach can also prevent overlooking potential new data uses that align with the user's original intent. The taxonomy of data reuse proposed by Custers and Ur{\v{s}}i{\v{c}}~\citep{custers2016big} could serve as a starting point to contextualized repurposing implementations.

\subsection{A temporal perspective}
User intentions can evolve over time, and with the ever-changing technological landscape, what was once feasible might now be drastically different. This brings up potential challenges when repurposing data: if too much time has elapsed since the initial consent was given, it might no longer be appropriate to reuse the data for even a seemingly compatible new purpose. Additionally, individual perceptions regarding the sensitivity and nature of personal data can shift as time goes on. The continuous reassessment of compatibility proposed in Sec.~\ref{sec:continuous-assessments} might be an appropriate method to address these temporal concerns.

\subsection{An economic perspective}
The problem of data repurposing lends itself to an economic analysis for several reasons. First, there might be a tension between service providers, who have an economic incentive in specifying purposes as broadly as possible, and regulators, who might argue for narrow definitions of purposes and their compatibility. Furthermore, studying to what extent users have the possibility to reveal their true preferences for repurposing policies through a behavioral analysis, might provide both descriptive guidelines regarding these policies, as well as an evaluation of the manipulative potential of the current mechanisms. Last but not least, an antitrust angle might reveal to what extent users have the ability to migrate between services should they be unhappy with the repurposing practices.

\section{Conclusions}
In this paper, I have analyzed the legal mechanism of purpose compatibility assessment from a computational perspective. I first sought to identify computational tools that could automate the compatibility criteria. Although complete automation was not feasible, I have emphasized specific aspects of the first three criteria (links between purposes, context of processing and reasonable expectations of users, and the nature of personal data) that could be quantified. 

The empirical portion of this paper focused on a case study of repurposing within recommender systems, revealing additional challenges of automated measurements. These challenges  included the difficulty of generalizing the measurement approaches to different computational scenarios, the availability of suitable data, and data biases. Moreover, I have discussed the necessity for research that connects the measurement of purpose links with the concrete experiences of data subjects.

Subsequently, I highlighted the ambiguity surrounding the definition of repurposing. I argued that aside from evident changes in the computational task, changes in the algorithms, data distributions, and the user base, should also be considered as repurposing and subjected to compatibility assessment tests. Consequently, I proposed that compatibility assessment should be an ongoing process facilitated by system monitoring with a comprehensive set of metrics.

Considering that compatibility must be established between two specific purposes, in the final analysis in this paper, I examined whether the current common framing of data processing purposes allows for meaningful repurposing. I identified several issues with purpose formulations, particularly the problem of excessively broad definitions that encompass most system functionalities. Consequently, despite the nuanced nature of repurposing from a computational standpoint, in practice, most of this nuance could be disregarded. Due to purpose formulations being the result of ``skilled legal drafting''~\citep[p. 6]{Finck_Biega_2021}, numerous types of data processing trivially fall under a limited number of purposes. However, given that data repurposing can lead to technological innovation and reduce the consent burden placed on data subjects, striking the right balance in purpose formulations will be crucial.

At a meta-level, the contributions of this paper are two-fold. Firstly, our work contributes to the ongoing discourse surrounding the automation of data protection compliance. Predicting a priori which data protection provisions can be fully or partially automated is challenging. Our current research, through its in-depth exploration of repurposing within a specific computational task, sheds light on the intricacies and viability of automating compatibility assessments.

Secondly, this paper unveils the potential for refining operational guidelines to compatibility assessment. Refined guidelines could not only bolster enforcement but also improve data engineering practices. Our research posits that compatibility assessments should ideally be a collaborative endeavor between computational and data protection specialists, involving incorporation of computational evidence directly into the repurposing decision-making processes.

In essence, our paper demonstrates a 'middle-way' strategy for compatibility assessments: a semi-automated compliance approach, positioning the provision as a constructive practice that strikes a balance between a de facto ban and an ineffective standard.

\bibliography{main} 

%
%

\begin{2Addresses}
{Asia J. Biega\newline
Max Planck Institute for Security and Privacy\newline
Universitaetstr. 140\newline
44799 Bochum, Germany\newline
asia.biega@mpi-sp.org}
&
{}
\end{2Addresses}

\end{document}